\documentclass[lettersize,conference]{IEEEtran}
\usepackage{amsmath,amsfonts}
\usepackage{algorithmic}
\usepackage{algorithm}
\usepackage{array}
\usepackage[caption=false,font=normalsize,labelfont=sf,textfont=sf]{subfig}
\usepackage{textcomp}
\usepackage{stfloats}
\usepackage{url}
\usepackage{verbatim}
\usepackage{graphicx}
\usepackage{cite}
\usepackage{bm}
\usepackage{xspace}
\newcommand{\ie}{i.e.,\xspace}

\begin{document}

\title{Making Locality-aware GEMM Compatible with Page-Granularity Placement on Chiplet GPUs}

\author{
    \IEEEauthorblockN{
        Euijun Chung,
        Jae Hyung Ju,
        Hyesoon Kim
    }

    \IEEEauthorblockA{
        \emph{
            \{euijun, jhju\}@gatech.edu, hyesoon@cc.gatech.edu
        }
    }

    \IEEEauthorblockA{
        Georgia Institute of Technology
    }
}

\markboth{IEEE Computer Architecture Letters,~Vol.~XX, No.~X, May~YYYY}%
{Chung \MakeLowercase{\textit{et al.}}: Making Locality-aware GEMM Compatible with Page-Granularity Placement on Chiplet GPUs}


\maketitle
\begin{abstract}
Multi-chiplet GPUs scale compute throughput and high-bandwidth memory (HBM) capacity, but their non-uniform memory system makes locality between chiplets and their data critical to the GPU's performance and energy efficiency.
Locality-aware scheduling and data placement identify which data should reside near each chiplet.
However, in general matrix multiplication (GEMM), locality-aware data placement often becomes incompatible with a fixed page-granularity data interleaving, since the optimal granularity for mapping data across chiplets varies widely across workloads.
We propose \emph{Chiplet-Contiguous Layout}, a global memory layout that stores chiplet-local data contiguously.
Chiplet-Contiguous Layout enables locality-aware placement compatible with page-granularity placement across diverse large language model (LLM) GEMM shapes, without changes to the operating system or hardware.
On representative LLM inference and training GEMMs from Qwen 3 30B and Llama 3.1 70B, Chiplet-Contiguous Layout on average reduces remote HBM traffic by 13.0$\times$ on Qwen and 20.7$\times$ on Llama over 4\,KB interleaving, and by 3.3$\times$ and 3.7$\times$ over coarse locality-aware placement.
\end{abstract}


\section{Introduction}

\begin{figure}
    \centering
    \vspace{-0.05in}
    \includegraphics[width=0.9\linewidth]{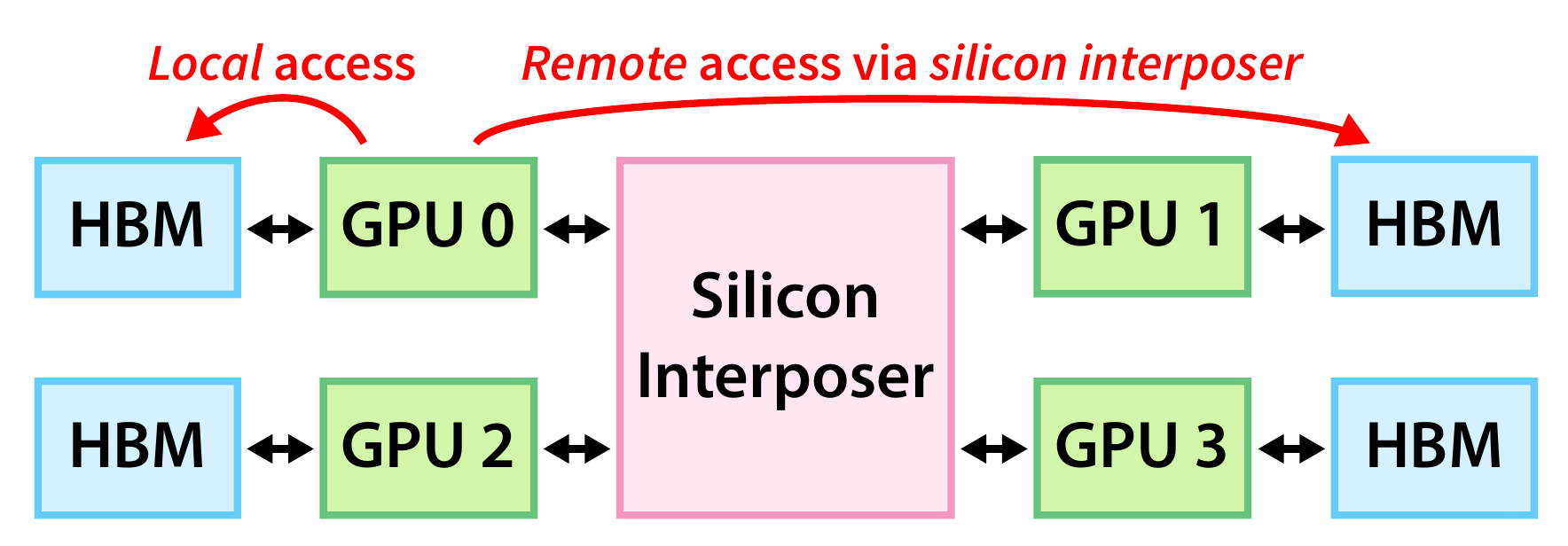}
    \vspace{-0.15in}
    \caption{Memory access on a 4-chiplet GPU. Remote access via a silicon interposer increases latency, power, and inter-chiplet bandwidth consumption.}
    \label{fig:chiplet}
    \vspace{-0.15in}
\end{figure}

\IEEEPARstart{M}{ulti-chiplet} GPUs are becoming an important design point for scaling GPU compute throughput and memory capacity beyond the reticle size limit.
By integrating multiple GPU chiplets and memory stacks into a single package via a silicon interposer, the system provides high aggregate bandwidth and capacity while improving yield and package-level scalability~\cite{arunkumar2017mcm, amd_cdna3_architecture_2025, nvidia_blackwell_architecture_2024}.
A multi-chiplet GPU may be exposed to programmers as a single logical GPU, but its memory system forms multiple non-uniform memory access (NUMA) regions within the package.
As shown in Figure~\ref{fig:chiplet}, a memory request generated by one GPU chiplet may be served either by a nearby HBM stack, \ie \textit{local} access, or by a remote HBM stack through the on-package interconnect, \ie \textit{remote} access.
Remote accesses consume shared inter-chiplet bandwidth and can increase latency and energy consumption compared to local accesses~\cite{arunkumar2017mcm, zhang2023characterizing}.
Therefore, improving memory locality is a central challenge in multi-chiplet GPU systems.

Prior work reduces remote traffic with locality-aware data placement and Cooperative Thread Array (CTA) scheduling by co-locating CTAs with the data regions they access~\cite{arunkumar2017mcm, kim2018coda, khairy2020locality}.
GEMM (general matrix multiplication) is a key operation for GPUs, and is a natural target for such co-location as its tiled execution exposes predictable CTA-to-operand affinity, as illustrated in Figure~\ref{fig:gemm}.
The optimal strategy for mapping matrix partitions onto chiplets varies with the GEMM shape and operand layout.
Across the forward and backward pass of feed-forward networks (FFNs) in LLMs, we find that for 67\% of the representative GEMMs in LLM training, remote HBM traffic is minimized when each operand is finely interleaved across chiplets along its contiguous dimension; also known as \textit{fine-grained data interleaving}~\cite{kim2018coda, khairy2020locality}.

However, conventional global memory layouts make it difficult to achieve this placement goal of fine-grained interleaving. 
The memory system assigns data to chiplets via \emph{page-granularity placement}: each fixed-size region, either a virtual memory page assigned by the operating system (OS), or a hardware-interleaved fixed-address block (e.g., 4\,KB), is mapped to a chiplet. 
Since the chiplet-local regions vary across GEMMs and are typically finer-grained than a single page (Figure~\ref{fig:problem}(a)), no virtual-to-physical mapping at page-granularity can place all chiplet-local data on its chiplet.

To this end, we propose \emph{Chiplet-Contiguous Layout} (CCL), a global memory layout for matrices that ensures data consumed by each chiplet is contiguous.
By making chiplet-local regions contiguous, CCL achieves optimal locality-aware data placement at page-granularity, requiring neither GEMM-dependent interleaving granularities in the memory system nor changes to the OS or hardware. 
Across 36 representative LLM GEMMs from Qwen 3 30B and Llama 3.1 70B, CCL reduces mean remote HBM traffic by 13.0$\times$ on Qwen and 20.7$\times$ on Llama over 4\,KB round-robin interleaving, and by 3.3$\times$ and 3.7$\times$ over coarse-grained locality-aware data placement.

\section{Background and Motivation}
\label{sec:bg}

\subsection{Data Placement Granularity in Chiplet GPUs}
\label{sec:bg_chiplet}

Locality-aware multi-chiplet GPU execution co-locates CTAs with the data regions they access, using compiler analysis, runtime profiling, or heuristics to infer CTA-to-data affinity~\cite{arunkumar2017mcm, kim2018coda, khairy2020locality, park2025leveraging}.
However, the memory system exposes placement only at the page-granularity, where each placement unit must be assigned to a single chiplet.
Thus, page-based placement is effective only when the data that should be local to one chiplet is contiguous and aligned at page-granularity.
A separate mechanism is hardware-level memory interleaving, where consecutive address regions are distributed across HBM stacks to balance bandwidth~\cite{amd_cdna3_architecture_2025}.
Because this mapping is fixed by address bits rather than CTA-to-data affinity, it is not locality-aware, but provides a useful baseline for traffic distribution without application-specific placement.

\begin{figure}[t]
    \centering
    \includegraphics[width=1.0\linewidth]{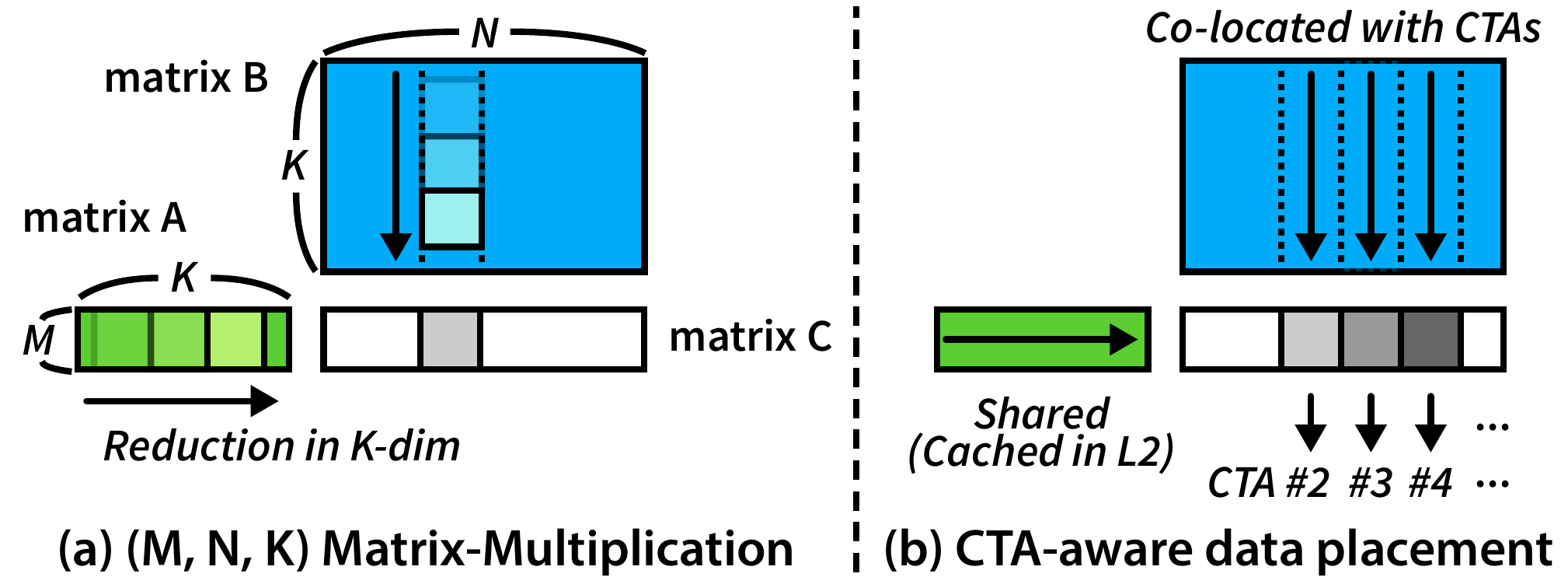}
    \vspace{-0.25in}
    \caption{(a) Tiled matrix-multiply. (b) Co-locating data with CTAs. When matrix $\mathbf{A}$ tiles are cached in each GPU chiplet's L2 cache, placing each tile column of matrix $\mathbf{B}$ near the CTAs that consume it allows the remaining HBM accesses to be served locally.}
    \label{fig:gemm}
    \vspace{-0.1in}
\end{figure}

\subsection{GEMM Locality and the Misalignment Problem}
\label{sec:bg_gemm}

We consider the canonical GEMM form $\mathbf{C}=\mathbf{A}\mathbf{B}$, where $\mathbf{A}\in\mathbb{R}^{M\times K}$, $\mathbf{B}\in\mathbb{R}^{K\times N}$, and $\mathbf{C}\in\mathbb{R}^{M\times N}$; the BLAS extension to $\alpha\mathbf{A}\mathbf{B}+\beta\mathbf{C}$ is straightforward, since the scalars do not affect placement and the $\mathbf{C}$ read shares the same partitioning as the output write.
Shown in Figure~\ref{fig:gemm}(a), GPU GEMM kernels decompose $\mathbf{C}$ into output tiles, with each CTA computing one tile and streaming the corresponding $\mathbf{A}$ and $\mathbf{B}$ tiles along $K$-dimension.
This tiled structure creates predictable CTA-to-data affinity: CTAs in the same output row reuse the same $\mathbf{A}$ tiles, while CTAs in different output columns consume different $\mathbf{B}$ tile columns.
As shown in Figure~\ref{fig:gemm}(b), a locality-aware scheduler can exploit this structure by placing operand regions near the chiplets that execute the corresponding CTAs~\cite{khairy2020locality}.

\begin{figure}[t]
    \centering
    \includegraphics[width=1.0\linewidth]{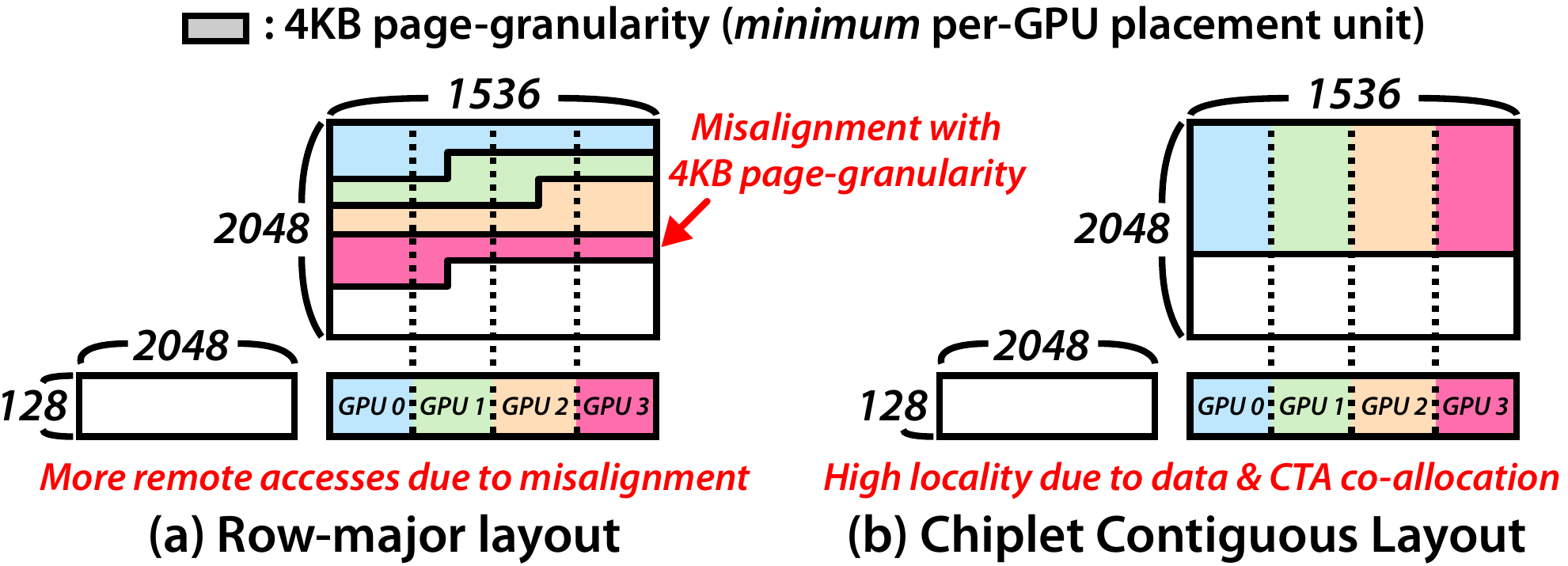}
    \vspace{-0.3in}
    \caption{Page-granularity placement on a Qwen 30B fused up/gate operand, BF16, 4 chiplets. (a) Row-major layout: chiplet ownership does not align with 4\,KB boundaries, so a single page contains data destined for multiple chiplets and forces remote accesses. (b) Chiplet-Contiguous Layout: per-chiplet strips are contiguous and aligned with page boundaries, enabling local accesses.}
    \label{fig:problem}
    \vspace{-0.15in}
\end{figure}

In GEMMs, the required placement granularity depends on the matrix dimensions and tensor layout.
As shown in Figure~\ref{fig:gemm}, placing different column regions of $\mathbf{B}$ on different chiplets is \textit{coarse-grained} when $\mathbf{B}$ is column-major, because each column region is contiguous in memory.
In row-major layout, however, the same column region appears as a small slice within every row, so achieving the same locality target requires splitting each row of $\mathbf{B}$ across chiplets, i.e., \emph{fine-grained interleaving}. 
For a row width of $N$ elements distributed across $G$ chiplets, this requires interleaving rows at a GEMM-dependent granularity of $N/G$ elements.

However, a \textit{misalignment problem} arises, where the data that should be chiplet-local is \textit{misaligned} with page-granularity boundaries (Figure~\ref{fig:problem}(a)).
Each CTA is then forced to fetch parts of its operand rows from remote chiplets, increasing the number of remote accesses via the silicon interposer.

This \textit{misalignment problem} arises because the required size of row slices (i.e., $1538/4=384$ elements in Figure~\ref{fig:problem}(a)) is not a multiple of the memory system's data placement granularity.
For BF16 data and four chiplets, row widths of 768, 4{,}096, and 14{,}336 elements in popular LLMs correspond to \emph{fine-grained data interleaving} with row-slices of 384\,B, 2\,KB, and 7{,}168\,B, respectively.
These matrix widths, deriving from LLM's hidden and intermediate dimensions, rarely aligns with fixed page sizes or hardware-interleaving granularities such as 4\,KB, 64\,KB, or 2\,MB, so a placement unit may contain data that should be placed near multiple chiplets, leaving some accesses remote.
Thus, the mismatch between GEMM dimensions and placement granularity limits the locality a placement policy can achieve.
This is a challenging problem to solve with the current global memory layout because GEMM dimensions vary across attention projections, feed-forward layers, and expert multi-layer perceptrons (MLPs).

\section{Chiplet-Contiguous Layout}
\label{sec:ccl}

\begin{figure}
    \centering
    \includegraphics[width=1.0\linewidth]{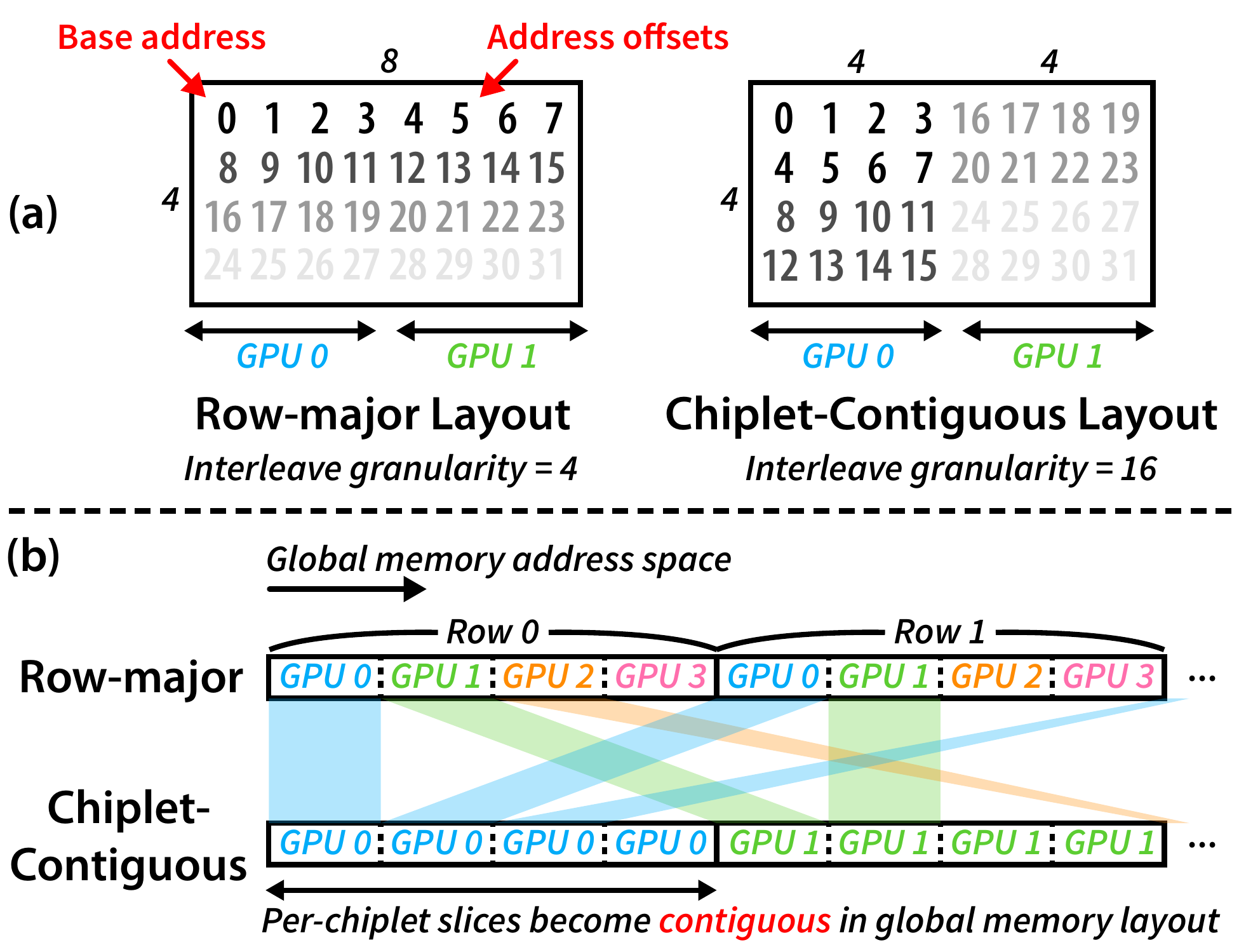}
    \vspace{-0.25in}
    \caption{Chiplet-Contiguous Layout (CCL) changes the global memory layout of a matrix. (a) Row-major layout makes rows contiguous, while CCL makes per-chiplet strips contiguous. (b) Row-major layout alternates chiplet slices within each row, whereas CCL groups slices by chiplet.}
    \label{fig:ccl}
    \vspace{-0.15in}
\end{figure}

\subsection{Remapping Row Interleaving into Contiguous Regions}
\label{sec:ccl_idea}

Our key insight is to change the tensor's global memory layout so that the data is expressed as contiguous chiplet-local regions, rather than irregular row slices.
Specifically, $\mathbf{C}$ is partitioned along chiplets, each CTA is scheduled on the chiplet owning its output tile, and our layout aligns $\mathbf{A}$ and $\mathbf{B}$ so that the same chiplet holds the regions consumed by that CTA, as shown in Figure~\ref{fig:problem}(b).
This layout eliminates the misalignment problem for most GEMM dimensions, while preserving the memory system's fixed page-granularity.

Figure~\ref{fig:ccl}(a) illustrates the difference between row-major and Chiplet-Contiguous Layout (CCL).
Both represent the same logical matrix and the same per-chiplet column slices.
In row-major layout, each row is stored contiguously, so the global memory order repeatedly alternates among slices for different chiplets within every row.
For example, with two chiplets, each row's slices for both chiplets are interleaved in memory, requiring an interleave granularity of 4 elements.
In contrast, CCL first stores all row slices consumed by chiplet 0, then all row slices consumed by chiplet 1, and so on.
Thus, the logical matrix and the locality-aware data placement are unchanged, but CCL enables an interleave granularity of 16 elements.

CCL makes 4\,KB interleaving practical.
Because each strip is contiguous, it can be aligned to the 4\,KB interleaving granularity instead of requiring GEMM-dependent row-slice interleaving, as shown in Figure~\ref{fig:ccl}(b).

\subsection{Layout Mapping and Page-Granularity Placement}
\label{sec:ccl_mapping}

\begin{figure}
    \centering
    \includegraphics[width=1.0\linewidth]{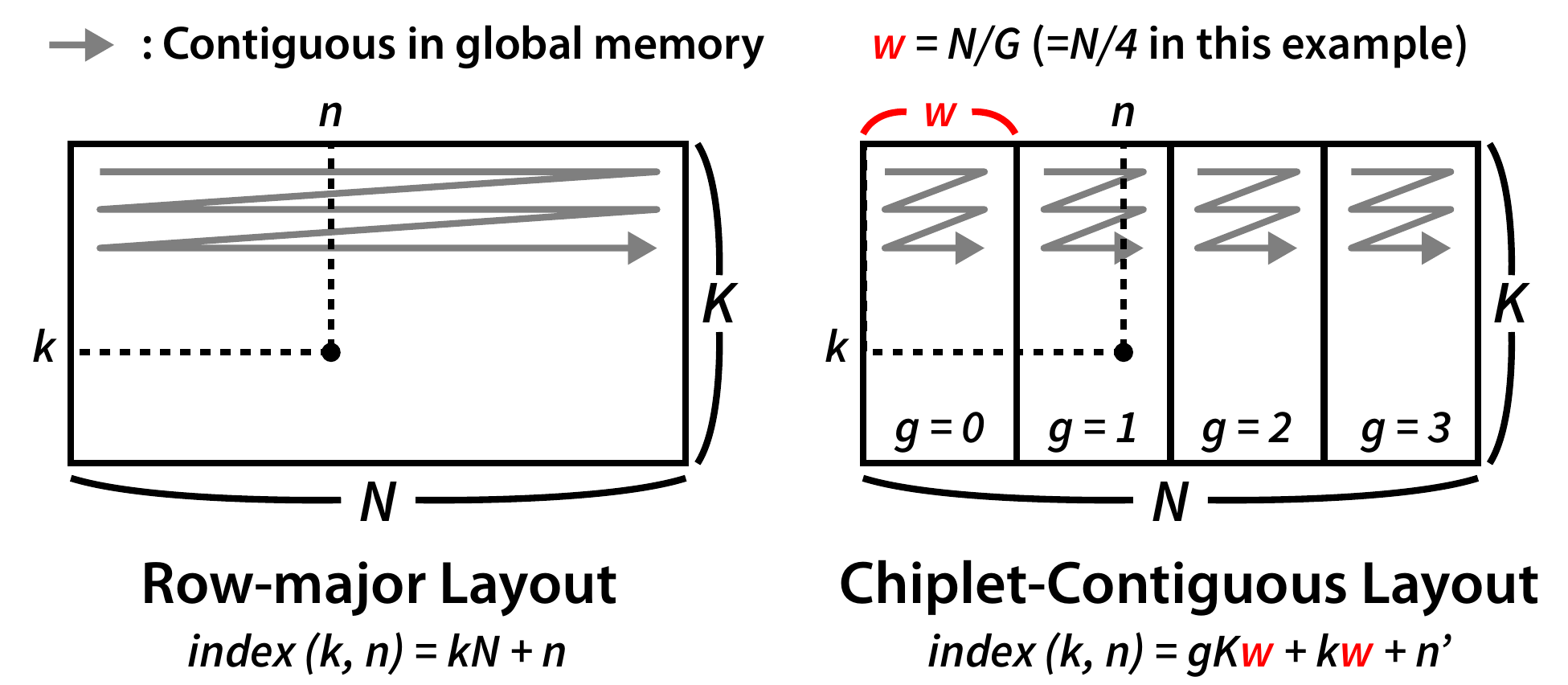}
    \vspace{-0.25in}
    \caption{Row-major and Chiplet-Contiguous address mapping for element $(k,n)$ in a $K\times N$ matrix. Chiplet-Contiguous Layout decomposes $n$ into a chiplet index $g$ and local offset $n'$. $w$ denotes the per-chiplet matrix width.}
    \vspace{-0.15in}
    \label{fig:mapping}
\end{figure}

We describe CCL's address mapping in Figure~\ref{fig:mapping}, using the $\mathbf{B}$ operand in Figure~\ref{fig:gemm}.
Let $\mathbf{B} \in \mathbb{R}^{K \times N}$ be distributed across $G$ chiplets along the $N$ dimension.
We define the per-chiplet width as $w=N/G$ and decompose each column index $n \in \{0, \cdots, N-1\}$ into a chiplet index $g \in \{0, \cdots, G-1\}$ and an intra-chiplet column index $n' \in \{0, \cdots, N/G - 1\}$:
\begin{equation}
    g = \left\lfloor \frac{n}{w} \right\rfloor, \qquad
    n' = n \bmod w .
\end{equation}
Here, $g$ identifies the chiplet-local strip that contains column $n$, and $n'$ identifies the column position within that strip.

In conventional row-major layout, element $\mathbf{B}[k,n]$ is stored at element index:
\begin{equation}
    \text{index}_{\text{row}}(k,n) = k \cdot N + n .
    \label{eq:row}
\end{equation}
This layout makes each row contiguous, but it also causes chiplet-local slices to repeat within every row.
CCL instead stores data in chiplet-strip order:
\begin{equation}
    \text{index}_{\text{CCL}}(k,n)
    =
    g \cdot K \cdot w + k \cdot w + n' .
    \label{eq:ccl}
\end{equation}
All elements with the same chiplet index $g$ therefore occupy one contiguous strip of $K \cdot w$ elements.
Thus, data that row-major layout spreads across many per-row slices becomes a single contiguous region for each chiplet, making 4\,KB interleaving locality-friendly.
Instead of matching the per-row slice granularity of $w$ elements, the memory system can align the full chiplet-local strip of size $K \cdot w$ elements to 4\,KB boundaries without requiring irregular per-row interleaving.

\subsection{Software Support}
\label{sec:ccl_software}

Kernel software implements CCL's address mapping by modifying the global memory layout descriptor.
Frameworks such as CUTLASS/CuTe already separate logical tensor coordinates from physical layouts through layout abstractions~\cite{nvidia2026cutlass}, so we express CCL as a custom global memory layout rather than a new kernel algorithm.
For example, for a matrix $\mathbf{B}\in\mathbb{R}^{K\times N}$ distributed across $G$ chiplets, we implement CCL by reshaping the logical view from $(K,N)$ to $(K,G,N/G)$ with strides $(w, K\cdot w, 1)$, so that the $G$ mode is the outermost.
CCL applies to all matrix operands ($\mathbf{A}$, $\mathbf{B}$, $\mathbf{C}$) stored in global memory: weights adopt CCL layout offline or at model-load time and are reused across many GEMMs, while activations are produced directly by upstream kernels in CCL layout or repacked when profitable.
Thus, GEMM computation is unchanged: CTAs compute the same logical output tiles, but each operand uses the CCL address mapping in Eq.~\eqref{eq:ccl} instead of the row-major mapping in Eq.~\eqref{eq:row}.
The resulting per-chiplet strips are compatible with both page-granularity placement and fixed-granularity HW interleaving; we use 4\,KB round-robin interleaving in our evaluation, where chiplet assignment is determined by address bits, and CCL aligns this address-driven mapping with the desired chiplet locality.
Translating between Eq.~\eqref{eq:row} and Eq.~\eqref{eq:ccl} adds only a few ALU operations per access, which are fully overlapped with HBM latency.

\section{Evaluation}
\label{sec:evaluation}

\subsection{Methodology}
\label{sec:eval_methodology}

To study how layout affects memory locality, we measure remote HBM traffic for representative LLM GEMMs under different data layouts and memory-mapping granularities.

\noindent \textbf{Simulation framework.} We use a custom tile-level GEMM locality simulator that models CTA execution, per-chiplet L2 caches, and HBM accesses.
Each CTA computes one output tile and issues operand accesses according to the GEMM shape and CTA traversal order.
The simulator classifies L2 misses as local or remote HBM accesses based on the data layout and memory-mapping policy.
Since a GEMM accesses each operand region in a fixed, balanced pattern across chiplets, dynamically migrating pages only shifts remote accesses to another rather than eliminating them. Therefore, we do not model page migration.

\noindent \textbf{Hardware Configuration.}
We use an MI300X-like configuration with four memory-locality domains (hereafter chiplets), each comprising two Accelerator Compute Dies (XCDs), for a total of 304 Compute Units (CUs). Each domain has an 8\,MB private L2 cache with 128\,B cache lines and 16-way associativity. GEMM kernels use $128\times128$ output tiles.

\begin{figure*}[t]
    \centering
    \vspace{-0.03in}
    \includegraphics[width=0.95\linewidth]{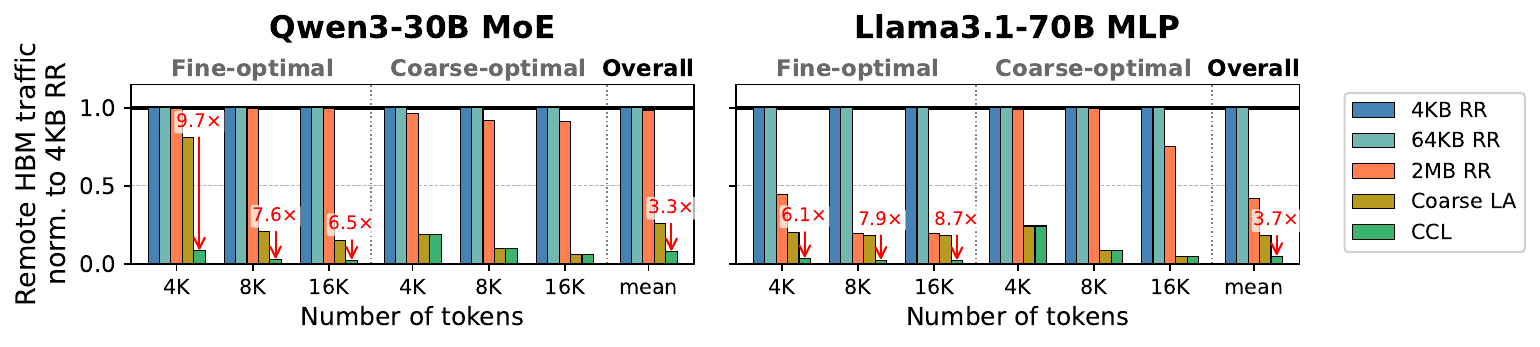}
    \vspace{-0.2in}
    \caption{Remote HBM traffic normalized to 4\,KB round-robin (RR) interleaving, for Qwen 3 30B (left) and Llama 3.1 70B (right). GEMMs are grouped into two groups (fine-/coarse-optimal) depending on their optimal locality-aware data placement strategy. Each swept over token counts of 4K, 8K, and 16K.}
    \label{fig:eval}
    \vspace{-0.15in}
\end{figure*}

\noindent \textbf{Workloads.}
We evaluate the gate/up and down projection GEMMs of the Qwen 3 30B (mixture-of-experts, MoE) and Llama 3.1 70B feed-forward networks (FFNs), in both the forward and backward pass. 
We assume each FFN, including the Qwen MoE backward, executes on a single GPU. 
To cover realistic per-GPU shapes for LLM training, we sweep three token counts (batch size $\times$ sequence length $=$ 4K, 8K, 16K), yielding 36 BF16 GEMMs in total. 
We group these GEMMs by their locality-optimal granularity: \emph{coarse-optimal} GEMMs, where partitioning every operand into large contiguous blocks minimizes remote HBM traffic, and \emph{fine-optimal} GEMMs, where only fine-grained interleaving does so; 24 of the 36 (67\%) GEMMs are \emph{fine-optimal}.


\noindent \textbf{Baselines.}
We compare CCL against fixed-granularity round-robin interleaving at 4\,KB, 64\,KB, and 2\,MB, which assigns consecutive fixed-size regions to chiplets in address order. 
This models the hardware interleaving used in AMD MI300X's single-logical-GPU (SPX) mode~\cite{amd_cdna3_architecture_2025}.
We also include coarse locality-aware placement (Coarse LA)~\cite{khairy2020locality}, which applies coarse-grained partitioning to every GEMM, splitting each matrix into $G$ large contiguous blocks per chiplet. 
CCL instead selects coarse- or fine-grained partitioning per GEMM: for fine-optimal GEMMs, it stores each chiplet's data contiguously in global memory so that plain 4\,KB round-robin interleaving achieves the fine-grained locality.
We sweep CTA traversal and output-partition choices and report the configuration with the lowest remote HBM traffic.

\noindent \textbf{Metrics.} 
Remote HBM traffic reports remote operand reads that miss in L2, and remote output writes.
We normalize traffic to the 4\,KB round-robin baseline.
Since the absolute traffic varies by orders of magnitude across GEMMs, we report the geometric mean of remote-traffic ratios in Figure~\ref{fig:eval}.

\subsection{Interposer Traffic Results}
\label{sec:eval_results}

This experiment supports our primary claim that CCL reduces remote HBM traffic compared to fixed-granularity and Coarse LA placement across LLM GEMMs. Figure~\ref{fig:eval} shows that changing fixed-granularity interleaving from 4\,KB to 2\,MB does not consistently reduce remote HBM traffic, because the placement unit still does not match GEMM-dependent chiplet-local slices. 
Coarse LA reduces traffic on some GEMMs when partitioning aligns with the GEMM access pattern, i.e., on \emph{coarse-optimal} GEMMs, but on \emph{fine-optimal} GEMMs, it inflates remote HBM traffic by up to 16.5$\times$ over CCL on Qwen (17.0$\times$ on Llama). 
In contrast, CCL makes each chiplet's data contiguous before placement, so even 4\,KB round-robin interleaving achieves fine-grained locality. 
CCL reduces mean remote HBM traffic to 7.7\% of the 4\,KB round-robin baseline for Qwen and 4.8\% for Llama, corresponding to 13.0$\times$ and 20.7$\times$ reductions. 
Compared to Coarse LA, CCL further reduces it by 3.3$\times$ on Qwen and 3.7$\times$ on Llama. 
These results show that the primary limitation is the mismatch between conventional tensor layouts and the memory system's locality granularity.

\begin{figure}[t]
    \centering
    \vspace{-0.05in}
    \includegraphics[width=0.9\linewidth]{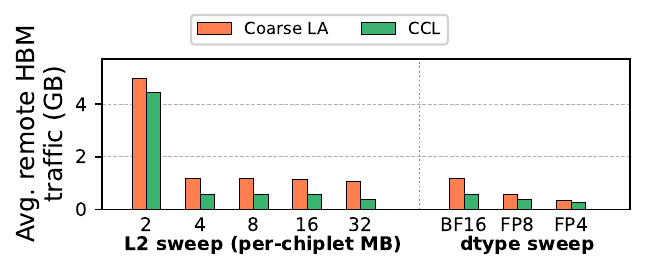}
    \vspace{-0.2in}
    \caption{Sensitivity of remote HBM traffic to per-chiplet L2 capacity and data type. Average absolute traffic across GEMMs with 4K tokens is reported. }
    \label{fig:sweep}
    \vspace{-0.15in}
\end{figure}

We test whether CCL's locality advantage holds across per-chiplet L2 capacities and operand data types.
Figure~\ref{fig:sweep} shows that CCL remains below Coarse LA both when sweeping L2 capacity with BF16 fixed and when sweeping data type with the per-chiplet L2 fixed at 8\,MB.

\section{Conclusion}
\label{sec:conclusion}

Multi-chiplet GPUs require data locality, but achieving it for GEMMs can require GEMM-dependent placement finer than a page, which conventional layouts cannot express. 
CCL groups data consumed by the same chiplet into contiguous global memory regions compatible with 4\,KB interleaving, achieving fine-grained locality without memory-system changes.
Evaluated on LLM GEMMs, CCL reduces remote HBM traffic over 4\,KB interleaving and coarse locality-aware placement. 

\clearpage

\bibliographystyle{IEEEtran}
\bibliography{refs}

@inproceedings{arunkumar2017mcm,
  title={MCM-GPU: Multi-chip-module GPUs for continued performance scalability},
  author={Arunkumar, Akhil and Bolotin, Evgeny and Cho, Benjamin and Milic, Ugljesa and Ebrahimi, Eiman and Villa, Oreste and Jaleel, Aamer and Wu, Carole-Jean and Nellans, David},
  booktitle={2017 ACM/IEEE 44th Annual International Symposium on Computer Architecture (ISCA)},
  pages={320--332},
  year={2017},
  organization={IEEE}
}

@techreport{amd_cdna3_architecture_2025,
  title       = {{AMD CDNA\texttrademark{} 3 Architecture: The All-New AMD GPU Architecture for the Modern Era of HPC and AI}},
  institution = {{Advanced Micro Devices, Inc.}},
  type        = {White Paper},
  year        = {2025}
}

@techreport{nvidia_blackwell_architecture_2024,
  title       = {{NVIDIA Blackwell Architecture Technical Brief: Powering the New Era of Generative AI and Accelerated Computing}},
  institution = {{NVIDIA Corporation}},
  type        = {Technical Brief},
  month       = mar,
  year        = {2024}
}

@article{kim2018coda,
  title={Coda: Enabling co-location of computation and data for multiple gpu systems},
  author={Kim, Hyojong and Hadidi, Ramyad and Nai, Lifeng and Kim, Hyesoon and Jayasena, Nuwan and Eckert, Yasuko and Kayiran, Onur and Loh, Gabriel},
  journal={ACM Transactions on Architecture and Code Optimization (TACO)},
  volume={15},
  number={3},
  pages={1--23},
  year={2018},
  publisher={ACM New York, NY, USA}
}

@inproceedings{khairy2020locality,
  title={Locality-centric data and threadblock management for massive GPUs},
  author={Khairy, Mahmoud and Nikiforov, Vadim and Nellans, David and Rogers, Timothy G},
  booktitle={2020 53rd Annual IEEE/ACM International Symposium on Microarchitecture (MICRO)},
  pages={1022--1036},
  year={2020},
  organization={IEEE}
}

@inproceedings{park2025leveraging,
  title={Leveraging Chiplet-Locality for Efficient Memory Mapping in Multi-Chip Module GPUs},
  author={Park, Junhyeok and Jang, Sungbin and Kwon, Osang and Lee, Yongho and Hong, Seokin},
  booktitle={Proceedings of the 58th IEEE/ACM International Symposium on Microarchitecture},
  pages={1040--1057},
  year={2025}
}

@misc{nvidia2026cutlass,
  title={NVIDIA CUTLASS Documentation},
  author={NVIDIA},
  howpublished={\url{https://docs.nvidia.com/cutlass/latest/}},
  year={2026},
}

@article{zhang2023characterizing,
  title={Characterizing multi-chip GPU data sharing},
  author={Zhang, Shiqing and Naderan-Tahan, Mahmood and Jahre, Magnus and Eeckhout, Lieven},
  journal={ACM Transactions on Architecture and Code Optimization},
  volume={20},
  number={4},
  pages={1--24},
  year={2023},
  publisher={ACM New York, NY}
}

\end{document}